
\input phyzzx

\def\dplus{=\hskip-5pt \raise 0.7pt\hbox{${}_\vert$}
^{\phantom 7}}
\def\dplusup{=\hskip-5.1pt \raise 5.4pt\hbox{${}_\vert$}
^{\phantom 7}}
\def\dplus{=\hskip-4.8pt \raise 0.7pt\hbox{${}_\vert$}
^{\phantom 7}}

\def\pmb#1{\setbox0=\hbox{#1} \kern-.025em\copy0\kern-\wd0
\kern0.05em\copy0\kern-\wd0 \kern-.025em\raise.0433em\box0}

\def\cH{{\cal H}}
\def\ubx{{\underline{x}}}
\def\uby{{\underline{y}}}

\font\mybb=msbm10 at 11pt

\def\bb#1{\hbox{\mybb#1}}

\def\bZ {\bb{Z}}
\def\bR {\bb{R}}
\def\rre {{\rm Re}}
\def\iim {{\rm Im}}

\def\bC {\bb{C}}

\def\e  {\epsilon}

\def\dd{d^\dagger}
\hfuzz 1cm

\REF\pt{G. Papadopoulos and P.K. Townsend,
{\it Compactification
of $D=11$ Supergravity on Spaces of Exceptional Holonomy},
Phys. Lett {\bf B357} (1995) 300: hep-th/9506150.}
\REF\hm{J. A. Harvey and
G. Moore, {\it  Superpotentials and Membrane Instantons}
 hep-th/9907026.}
\REF\candelas{P. Candelas and X. de la Ossa,
{\it Moduli Space of Calabi-Yau
Manifolds}, Nucl. Phys. {\bf B355} (1991) 455.}
\REF\pioline{H. Partouche and B. Pioline, {\it
Rolling Among $G_2$ Vacua},
JHEP 0103:005, 2001; hep-th/0011130.}
\REF\joyce{D.D. Joyce, {\it Compact Riemannian
7-Manifolds with
Holonomy $G_2$}, I, II, J. Diff. Geom.
{\bf 43} (1996) 291 and 329.}
\REF\malda{C.G. Callan, Jr. and J. Maldacena,
{\it D-Brane Approach
to Black Hole Quantum Mechanics}, Nucl. Phys.
{\bf B472} (1996) 591; hep-th/9602043.}
\REF\fks{S. Ferrara, R. Kallosh and A. Strominger,
{\it $N=2$ Extremal Black Holes},
Phys. Rev. {\bf D52} (1995) 5412; hep-th/9508072.}
\REF\moore{ G. Moore, {\it Attractors
and Arithmetic}, hep-th/9807056.}
\REF\hitchin{N. Hitchin, {\it The geometry of
 three-forms in six and seven dimensions}, math.DG/0010054.}
\REF\cg{ A. Comtet and G.W. Gibbons, {\it
Bogomolny Bounds for
Cosmic Strings}  Nucl.Phys. {\bf B299}:719, 1988.}
\REF\vy{ B. R. Greene, A. Shapere, C. Vafa, S-T
Yau, {\it Stringy Cosmic Strings and
Noncompact Calabi-Yau Manifolds}, Nucl.Phys.
{\bf B337}:1, 1990. }
\REF\gpb{G. Papadopoulos and P.K. Townsend,
{\it Intersecting M-branes},
Phys. Lett. {\bf B380} (1996) 273; hep-th/9603087.}
\REF\tseytlin{M. Cveti\v c and A.A. Tseytlin, {\it Non-extreme
Black Holes from Non-extreme Intersecting M-Branes},
Nucl. Phys. {\bf B478} (1996) 181; hep-th/9606033.}
\REF\cfgp{E. Cremmer, S. Ferrara, L. Girardello, A. Van
Proeyen, {\it Yang-Mills Theories with Local
Supersymmetry: Lagrangian,
Transformation Laws and Superhiggs Effect},
 Nucl.Phys. {\bf B212}:413,1983.}
\REF\bw{E. Witten, J. Bagger, {\it
Quantization of Newton's Constant in
Certain Supergravity Theories},
 Phys.Lett.{\bf B115}:202,1982. }
 \REF\wess{ J.A.  Bagger and J. Wess,
{\it Supersymmetry and Supergravity},
Princeton University Press, (1992).}
\REF\hklr{P. Howe, A. Karhede, U. Lindstr\"om  and M. Rocek,
{\it The Geometry of Duality}, Phys. Lett.
{\bf 168B} (1986) 98.}
\REF\csj{E. Cremmer, B. Julia and J. Scherk,
{\it Supergravity in Eleven-Dimensions}
Phys. Lett. {\bf B76} (1978) 403.}
\REF\gukov{ S. Gukov, {\it Solitons, Superpotentials and
Calibrations}, Nucl. Phys. {\bf B574} (2000) 169;
hep-th/9911011.}
\REF\raf{R. Hernandez, {\it Calibrated Geometries and Non
Perturbative  Superpotentials in M-Theory}, Eur. Phys. J.
{\bf C18} (2001) 619; hep-th/9912022.}
\REF\spen{B. S. Acharya and B. Spence, {\it Flux,
Supersymmetry and M Theory on 7-Manifolds}, hep-th/0007213.}
\REF\becker{K. Becker, {\it Compactifying M-theory
to Four-Dimensions}, hep-th/0010282.}
\REF\hl{R. Harvey and H.B. Lawson, {\it Calibrated Geometries},
 Acta Mathematica {\bf 148} (1982) 47.}
\REF\strombeck{K. Becker, M. Becker and A. Strominger,
{\it Five-Branes, Membranes and Non-Perturbative String Theory},
Nucl. Phys. {\bf B456} (1995) 130; hep-th/9507158.}
\REF\kallosh{E. Bergshoeff, R. Kallosh, T. Ortin
and G. Papadopoulos,
{\it Kappa Symmetry, Supersymmetry and Intersecting Branes},
Nucl. Phys. {\bf B502} (1997) 149; hep-th/9705040.}
\REF\macl{R.C. McLean, {\it Deformations of
 Calibrated Submanifolds},
Commun. in Analysis and Geometry, {\bf 6} (1998) 705.}
\REF\joyceb{D.D. Joyce, {\it Compact Manifolds
 with Special Holonomy}
Oxford University Press, (2000).}
\REF\gwggp{G.W. Gibbons and G. Papadopoulos, {\it Calibrations and
Intersecting Branes}, Commun. Math. Phys. {\bf 202}
(1999) 593; hep-th/9803163.}
\REF\west{J. P. Gauntlett, N. D. Lambert and P.C. West,
{\it Branes and Calibrated Geometries}, Commun. Math. Phys.
{\bf 202} (1999) 571; hep-th/9803216.}
\REF\jggp {J. Gutowski and G. Papadopoulos,
{\it Magnetic Cosmic Strings of N=1, D = 4 Supergravity
with Cosmological Constant}: hep-th/0102165.}
\REF\wipo{J. Polchinski, E. Witten
{\it  Evidence for Heterotic - Type I String Duality},
 Nucl.Phys. {\bf B460}:525-540,1996;
 hep-th/9510169.}
  \REF\ptb{E. Bergshoeff, M. de Roo , M.B. Green,
G. Papadopoulos, P.K. Townsend,
 {\it Duality of Type II 7 Branes and 8 Branes},
  Nucl.Phys. {\bf B470}:113-135,1996;
 hep-th/9601150.}
 \REF\schoen{R. Schoen and S.T. Yau, {\it Lectures on Harmonic Maps}
Conference Proceedings and Lecture Notes in Geometry and Topology,
Vol II, International Press (1997).}
\REF\cveticd{M. Cveti\v c and H. H. Soleng,
 {\it Supergravity Domain Walls}, Phys. Rep. {\bf 282}
(1997) 159;
hep-th/9604090.}
\REF\shiraishi{ K. Shiraishi, {\it Moduli Space
 Metric for Maximally-Charged
Dilaton Black Holes}, Nucl. Phys. {\bf{B402}} (1993) 399.}


\Pubnum{ \vbox{ \hbox{}\hbox{} } }
\pubtype{}
\date{October, 2001}
\titlepage
\title{Moduli Spaces and Brane Solitons  for  M-Theory
Compactifications on Holonomy $G_2$ Manifolds}
\author{J. Gutowski }
\address{Department of Physics,  Queen Mary College,
Mile End \break
London, E1 4NS}
\andauthor{G. Papadopoulos}
\address{Department of Mathematics, King's
College London,
 Strand \break
London, WC2R 2LS}

\abstract { We investigate the local geometry on the
moduli space of $G_2$ structures
that arises in compactifications of M-theory on holonomy
$G_2$ manifolds. In particular, we
determine the homogeneity properties
of  couplings of the associated $N=1$, $D=4$ supergravity
under the scaling of  moduli space coordinates.
We then find some brane solitons of $N=1$, $D=4$
supergravity that are associated with wrapping
M-branes on cycles of the compact space. These
include cosmic strings and domain walls
that preserve $1/2$ of supersymmetry of the
four-dimensional theory, and non-supersymmetric
electrically and magnetically charged black
holes. The geometry of some of the
black holes is that of non-extreme M-brane
configurations reduced
to four-dimensions on a seven torus. }

\vskip 1.0 cm
\endpage
\pagenumber=2
\font\mybb=msbm10 at 12pt
\def\bb#1{\hbox{\mybb#1}}

\def\C{\mkern1mu\raise2.2pt\hbox{$\scriptscriptstyle|$}
\mkern-7mu{\rm C}}

\def\log{{\rm log}}

\def\dA{{\dot{A}}}
\def\dB{{\dot{B}}}
\def\ddA{{\ddot{A}}}
\def\ddB{{\ddot{B}}}

\def\l {\lambda}
\def\d{\delta}

\def\a{\alpha}

\def\t{\tau}
\def\m{\mu}

\def\g{\gamma}
\def\pd{\partial_}

\def\pu{\partial^}
\def\b{\beta}
\def\g{\gamma}
\def\k{\kappa}
\def\f{\phi}
\def\s{\sigma}
\def\e{\epsilon}

\def\cM {{\cal {M}}}


\chapter{Introduction}

Compactifications of M-theory  on
 manifolds with $G_2$ holonomy
provide a way  of constructing four-dimensional
effective theories
 which have a  realistic amount of
supersymmetry.
These   effective theories are
$N=1, D=4$ supergravities with
 field content which
is determined by the Betti numbers of the compact space.
In particular it has been shown that  the associated
$N=1, D=4$  supergravity has $b_2$ vector  and $b_3$
chiral multiplets [\pt].
Reducing $N=1, D=11$ supergravity on compact
$G_2$ holonomy manifolds, one can also determine
the couplings of the four-dimensional theory as a function
of the various moduli fields [\pt, \hm]. These couplings
are naturally interpreted in terms of the geometry
of the moduli space of $G_2$ structures in a
 similar way to that
of Calabi-Yau compactifications of string
 theory, see for example
[\candelas]. Many compactifications of M-theory on holonomy
$G_2$ manifolds have been investigated, see
for example [\pioline],
 using the examples of holonomy $G_2$ manifolds
constructed by
Joyce in [\joyce].

Some of the  solutions of D=4 and D=5
supergravity theories which arise from
compactifications of strings and M-theory,
 like black holes, strings and domain walls, can be
 associated with branes wrapped on the homology
cycles of the compact manifold.
This correspondence between solutions of lower dimensional
supergravity theories and ten- and eleven-dimensional
 brane configurations has been very fruitful, for example it
has led to the microscopic computation
of the black hole entropy for a certain class of
extreme black holes [\malda].
Another application is the use  of  the behaviour
 of the black hole solutions of $N=2, D=4$
supergravity theories [\fks]  to provide evidence for the
existence of calibrated representatives for certain
homology cycles of Calabi-Yau manifolds [\moore].

In this paper, we shall examine the couplings of the
$N=1, D=4$ supergravity theories that
arise from compactifications
of $N=1, D=11$ supergravity on compact
 holonomy $G_2$ manifolds.
In particular we shall show that  the components of the
 metric of the sigma model manifold, which is $T\cM$,
 of $N=1, D=4$ theory are homogeneous of degree $-2$
under the scaling of certain coordinates of the
moduli space $\cM$
of $G_2$ structures.
For this we shall use a  result obtained  by Hitchin
[\hitchin] that the volume
of the compact $G_2$ manifold is homogeneous of
degree $7/3$ under the scaling
of some  coordinates of $\cM$.
It turns out that the metric on the moduli space
of $G_2$ structures
 is invariant under this scaling transformation;
 the isometry group of the sigma model metric on $T\cM$ is
generated
by the same scaling transformation and the translations along the
fibres.
In addition,  we shall show that the K\"ahler potential
of the sigma model manifold can be expressed in terms
 of the logarithm of the volume
of the compact holonomy $G_2$ manifold, see also [\hm].
The couplings of the vector multiplets to the
scalars are linear in some natural complex coordinates on the
sigma model manifold.

Having established the homogeneity properties
of the couplings
of $N=1, D=4$ supergravity associated with
compactifications of
$N=1, D=11$ supergravity on holonomy $G_2$
manifolds, we shall
explore the various solutions of the four-dimensional theory
that arise by wrapping M-branes on the homology cycles
of the compact manifold.
We shall find that the $N=1, D=4$ supergravity
admits string solutions
which preserve $1/2$ of  supersymmetry. These are associated
with M5-branes wrapped on coassociative  cycles of the
 compact manifold.
The form of these string solutions is that of
cosmic string solutions
of [\cg, \vy]. However the string solutions associated
with $G_2$ compactifications have infinite tension because the
sigma model manifold $T\cM$ is non-compact.
We shall also describe the M-theory origin of
domain wall solutions of $N=1, D=4$
supergravity which preserve $1/2$ of the supersymmetry.
The Killing spinor equations for such domain walls are
given in an appendix.
Next we shall explore the electric and magnetic
 black hole solutions of $N=1, D=4$
supergravity that arise from wrapping M2-branes and M5-branes
on 2- and 5-cycles of $N$, respectively. We shall show that
such solutions are not supersymmetric as expected.
We shall mainly focus in the case where
the only non-vanishing modulus
field is that corresponding to the overall scale
of the moduli space coordinates. We shall call
such solutions \lq\lq dilatonic''; we justify
this terminology in an appendix.
We shall find a class of extreme dilatonic
  solutions of the $N=1, D=4$ theory which
have the same spacetime geometry   as
 two supersymmetric  orthogonally intersecting
M-branes, eg two M2-branes intersecting
on a $0$-brane or two M5-branes intersecting on a 3-brane [\gpb],
reduced to four-dimensions on seven-dimensional torus.
Such four-dimensional solutions  exhibit a naked singularity.
We shall also present  some  dilatonic black hole
solutions. These have the same spacetime geometry
as that  of two  non-extreme orthogonally
intersecting M-branes reduced to
four-dimensions again on a seven-dimensional torus
found in  [\tseytlin].

We remark that our dilatonic domain wall and black
 hole solutions
depend only on the homogeneity properties of the couplings
of the $N=1, D=4$ supergravity. So  they will
 remain solutions
of the effective theory of $G_2$ compactifications
 after perturbative or non-perturbative corrections to the
couplings are taken into account which  preserve
these homogeneity properties.

The organization of  this paper is as follows:
In section two,
we present the action, Killing spinor equations
  and the geometry associated with the
couplings  $N=1, D=4$ supergravity with scalar
and vector multiplets.
In section three, we give the couplings of $N=1, D=4$
supergravity that arise from the compactification
of $N=1, D=11$ supergravity on holonomy
$G_2$ manifolds. We then present two approaches in
the investigation of the geometry of the moduli
space of $G_2$ structures. One is based on the
K\"ahler geometry and the other is based on the
symplectic geometry. We also express the metric
on the $G_2$ moduli space, that arises
in these compactifications, in terms of the volume
of the compact holonomy $G_2$ manifold.
In section four, we summarize some of the results
on calibrating cycles
in holonomy $G_2$ manifolds. We also give the
number of  supersymmetries
preserved by M-brane probes wrapping such cycles.
In section five, we present our string
solutions of $N=1, D=4$
supergravity associated with $G_2$ compactification.
In section six, we describe various domain walls.
In section seven, we give various
black hole solutions associated with
M2- and M5-branes wrapped
on homology 2- and 5-cycles of the compact manifold.
Finally in the appendices, we give our
spinor conventions, analyze the Killing spinor
equations of $N=1, D=4$ supergravity in connection
to strings and domain walls that
 arise in $G_2$ compactifications,
and give the bosonic action that describes
  the dilatonic black hole
system.

\chapter{$N=1$ $D=4$ Supergravity}

\section{Supergravity Action and Killing Spinor Equations}

The geometric data that determine the
couplings of  $N=1$ supergravity in
four-dimensions with
$n$ vector and
$m$ chiral multiplets that we shall use in this paper
are the following:

\item{(i)} A K\"ahler-Hodge manifold $M$ of
complex dimension $m$ with K\"ahler
potential $K$.

\item{(ii)} A  vector bundle $E$ over $M$ of
rank $n$ for which its complexified  symmetric product
admits a holomorphic section $h$.

\item{(iii)} A locally defined holomorphic function $f$ on $M$.

\item{(iv)} Sigma model maps, $z$, from the four-dimensional
spacetime $\Sigma$ into the manifold $M$.

\item{(v)} A principal bundle $P$ on the
four-dimensional spacetime
$\Sigma$ with fibre the abelian group $U(1)^n$ such that
the pull back of $E$ with respect to $z$ is isomorphic to
$P\times_{U(1)^n} {\cal L}U(1)^n$, where ${\cal L}U(1)^n$
is the Lie algebra of $U(1)^n$.

Given these data, the bosonic part  of $N=1, D=4$
supergravity action [\cfgp, \bw, \wess ] is
$$
L=\sqrt{-g}\big[ {1\over2} R(g)-
{1\over4} {\rm Re}h_{ab} F^a_{MN} F^b{}^{MN} +{1\over4}
{\rm Im}h_{ab} F^a_{MN} {}^\star F^b{}^{MN}-
\g_{i\bar j}\partial_M z^i\partial^M z^{\bar
j}-V \big]
\eqn\sugract
$$
where
$$
V=e^K[\g^{i\bar j} D_if  D_{\bar j}{\bar f}- 3 |f|^2]
+{1\over2} D_a D^a\ ,
\eqn\poteq
$$
$$
F^a_{MN}=\partial_M A_N^a-\partial_N A_M^a\ ,
\eqn\gageq
$$
$$
D_i f=\partial_i f+\partial_i K f\ ,
\eqn\coveq
$$
$A_N^a$ are $U(1)$ (Maxwell) gauge potentials and the
$D_a$ are constants associated to a Fayet-Iliopoulos term.
We remark that the gauge indices $a, b=1, \dots, n$ are
raised
and lowered with ${\rm Re} h_{ab}$; $i,j=1, \dots, m$ and
$M,N=0,\dots,3$ are holomorphic sigma model manifold
and spacetime indices, respectively.
Clearly $h_{ab}$ are the gauge couplings,
$f$ is a superpotential and $M$ is the sigma model manifold.

The above action may also describe the coupling of
$\ell$ linear
multiplets to $N=1$, $D=4$ supergravity [\hklr].
This is because the two-form
gauge potentials of the
linear multiplets can be dualized
to scalars in four dimensions. The resulting
action depends only on the spacetime
derivatives of dual scalars. Therefore  it is
invariant under
$\bR^\ell$ acting on these scalars with constant shifts.

In what follows some of the solutions of
the $N=1, D=4$ supergravity that
we shall consider  will be supersymmetric.
To explore their properties, we  need
the Killing spinor equations of \sugract\ which are the
vanishing conditions of the supersymmetry transformations
of the fermions of the theory.
These are most  conveniently expressed in terms
of a real 4-component Majorana spinor $\e$ as
$$
2 \big( \pd{M} +{1 \over 4} \omega_{M
{\underline A} {\underline B}}
\Gamma^{{\underline A} {\underline B}} \big) \e
-\big(  {\rm Im} (K_i \partial_M z^i)
+e^{K \over 2} \big( {\rm Re} f - {\rm Im} f \Gamma^5 \big)
\Gamma_M \big) \e=0\ ,
\eqn\supa
$$
$$
\big( -{1 \over 2} F{}^a{}_{MN} \Gamma^{MN} +
\Gamma^5 D^a \big) \e=0
\eqn\supb
$$
and
$$
\eqalign{
\big( {\rm Re} (\partial_M z^i) -
\Gamma^5 {\rm Im} (\partial_M z^i) \big) \Gamma^M \e
-e^{K \over 2} \big( {\rm Re} (\g^{i {\bar j}}
D_{\bar j} {\bar f})
- \Gamma^5
{\rm Im} (\g^{i {\bar j}} D_{\bar j} {\bar f}) \big)
\e =0\ ,}
\eqn\supc
$$
where underlined indices ${\underline A}, \ {\underline B}$
 denote tangent frame
indices and $\Gamma^5 =
 \Gamma^{\underline{0}} \Gamma^{\underline{1}}
\Gamma^{\underline{2}}
\Gamma^{\underline{3}}$. For our spinor conventions
see the  appendix.

The field equations of the supergravity action \sugract\
 are the following:

\item{(1)} The Einstein equations are:
$$
\eqalign{
G_{MN} - \rre h_{ab} F{}^a{}_{ML} F{}^b{}_N{}^L
- 2\g_{i \bar j} \partial_{(M} z^i \partial_{N)} z^{\bar j}
\cr
+g_{MN} \big({1 \over 4} \rre h_{ab} F{}^a{}_{LP} F^{bLP}
+ \g_{i \bar j}
 \partial_L z^i \partial^L z^{\bar j} +V \big) =0\ .}
\eqn\einga
$$

\item{(2)} The  Maxwell field equations  are:

$$
\pd{M} \big[ \sqrt{-g} \big( \rre h_{ab} F^{bMN} -
 \iim h_{ab} {}^\star F^{b MN} \big) \big]\ .
\eqn\maxgb
$$

\item{(3)} The scalar equations; varying $z^\ell$
gives the equation
$$
\eqalign{
-{1 \over 8} \pd{\ell} h_{ab} F{}^a{}_{MN} F^{b MN} -
\pd{\ell}V - {i \over 8} \pd{\ell} h_{ab} F{}^a{}_{MN}
{}^\star F^{b MN}
\cr
+\g_{\ell \bar j}
\big( \nabla_M \partial^M z^{\bar j}
+ \Gamma{}^{\bar j}{}_{\bar i \bar k} \pd{M} z^{\bar i}
\partial^M z^{\bar k} \big)=0\ ,}
\eqn\scalgc
$$
where $\nabla_M$ is the covariant derivative with
respect to the Levi-Civita
connection of the spacetime metric and
$$
\pd{\ell} V = \pd{\ell} \big( e^K \g^{i \bar j} D_i f\big)
D_{\bar j} {\bar f}  -2 e^K {\bar f} D_\ell f
+{1 \over 2} \pd{\ell} (D_a D^a)\ .
\eqn\varipot
$$
Taking the conjugate of this equation, one
obtains the field equation for
$z^{\bar \ell}$.

\chapter{Supergravity Actions from $G_2$ Compactifications}

\section{Compactification Ansatz}

The bosonic part of the $D=11, N=1$ supergravity
Lagrangian [\csj] that we shall consider is
$$
{\cal L}= {1\over2} {\sqrt h} R- {1\over4} F\wedge
 \star F+{1\over12} C\wedge F\wedge F
\eqn\elevgr
$$
where $h$ is the eleven-dimensional metric, $F$ is the
4-form\foot{Our form conventions
are $\chi={1\over k!} \chi_{I_1 \dots I_k} dx^{I_1}
\wedge\dots \wedge dx^{I_k}$, $(\chi,
\psi)={1\over k!}
\chi_{I_1,\dots,I_k} \psi^{I_1,\dots I_k}$ and $(\chi,
\psi)=\chi\wedge *\psi$.} field strength and $C$ is
the 3-form gauge potential, $F=dC$.

To derive the  $N=1, D=4$ supergravity action that arises
from the compactification of $N=1, D=11$ supergravity
on a holonomy $G_2$ manifold $N$, we introduce a  basis
of harmonic forms $\{\phi_i; i=1, \dots, m=b_3\}$
in $H^3(N, \bR)$ and similarly a basis
 $\{\omega_a; a=1, \dots, n=b_2\}$ in $H^2(N, \bR)$.
Repeating the analysis in  [\pt, \hm], we write
the compactification  ansatz for the eleven-dimensional
metric $ds^2$ and the three-form
gauge potential $C$ as
$$
\eqalign{
ds^2&=g_{MN}(x) dx^M dx^N+ G_{I J}(y, s(x)) dy^I dy^J
\cr
C&=\sum_{a=1}^n A^a(x)\wedge \omega_a(y)+
\sum_{i=1}^m p^i(x)\wedge \phi_i(y)\ ,}
\eqn\thrfrm
$$
where $G_{IJ}$ is the metric on $N$ with holonomy $G_2$
 depending
on the real  coordinates  $\{s^i; i=1, \dots, b_3\}$ of the
moduli space $\cM$ of the $G_2$ structures which have been
promoted to four-dimensional scalar fields;
$I,J=1,\dots, {\rm dim}N$. In addition, $A^a$ and $p^i$
are the one-form gauge potentials
 and real scalars of the four-dimensional theory,
respectively.
The fields $g, A^a, s^i, p^i$ describe the
small fluctuations
of the $G_2$ manifold $N$ within the $N=1, D=11$ supergravity.
To solve the field equations of $N=1, D=11$
supergravity at the
linearized level, the basis of harmonic forms
 $\{\phi_i; i=1, \dots, n=b_3\}$
in $H^3(N, \bR)$ is chosen with respect
to the $G_{IJ}$ metric
and similarly for the basis $\{\omega_a; a=1, \dots, n=b_2\}$
in $H^2(N, \bR)$.
So far the coordinates $s^i$ on the moduli space
 $\cM$ have been
 chosen in an arbitrary way. However below for
the investigation of the
couplings of the $N=1, D=4$ supergravity theory a
special choice will be made.

The  compactification of $N=1, D=11$ supergravity
 on holonomy
$G_2$ manifolds preserves four real supercharges. So
it is expected that the four-dimensional action
that describes
the dynamics of the small fluctuations of such
 background will exhibit $N=1, D=4$ supersymmetry.
Some of the couplings of the four-dimensional
 effective action
 can be easily deduced
from the eleven dimensional supergravity action and
are as follows:
$$
\eqalign{
ds^2&=\gamma_{i\bar j}dz^i d\bar z^j=k_{ij}(s) ds^i ds^j
+m_{ij}(s) dp^i dp^j
\cr
m_{ij}(s)&={1\over 4 \int_N \sqrt{G}\, d^7y} \int_N\,
\sqrt{G}\, d^7y\, (\phi_i, \phi_j)
\cr
{\rm Re} h_{ab}(s)&={1\over2}\int_N \sqrt{G}\,
d^7y\,(\omega_a, \omega_b)={1\over2}\int
\omega_a\wedge *\omega_b=-{1\over2}\int_N
\omega_a\wedge
\omega_b\wedge \phi
\cr
{\rm Im} h_{ab}(p)&=-{1\over2} p^i
 \int_N \omega_a\wedge\omega_b\wedge \phi_i=
-{1\over2} p^i C_{iab}\ ,}
\eqn\coupl
$$
where $\phi$ is the parallel 3-form associated with
 the $G_2$
structure on $N$ and we have used $G$ to also
denote the determinant
of the metric $G_{IJ}$ with holonomy $G_2$.
In an adapted frame $\{e_1, \dots, e_7\}$
of the  $G_ 2$ structure the
3-form $\phi$ can be written as
$$
\phi= (e_1 \wedge e_2-e_3\wedge e_4)\wedge e_5
+ (e_1\wedge e_3-e_4\wedge e_2)\wedge e_6
+(e_1\wedge e_4-e_2\wedge e_3)\wedge e_7+e_5
\wedge e_6\wedge e_7\ .
\eqn\calfrm
$$
The last equality in the third equation in \coupl\
 can be established using
$G_2$ representation theory, see for example
 [\joyce, \joyceb] and next section.
We remark that the intersection numbers
 $C_{iab}$ are topological and so
they do not depend on the moduli space
coordinates $s$ of $G_2$ structures or the
choice of harmonic representatives.
In addition, we remark that the couplings in \coupl\
do {\sl  not} depend on the
harmonic representatives chosen for the basis
$\omega_a$ in $H^2(N, \bR)$ and
the basis $\phi_i$ in  $H^3(N, \bR)$.
To express the four-dimensional couplings
 as above, we have rescaled the
four-dimensional metric $g\rightarrow \Theta^{-1} g$
with the volume $\Theta$ of the
compact space in order  to bring
the $D=4$ action in the Einstein frame.
Observe that the sigma model metric
 $ds^2$ above is invariant under the
action of $\bR^{b_3}$ acting with constant shifts on $p$.

Since the $G_2$ compactifications preserve four
real supercharges,
the effective theory has $N=1, D=4$ supersymmetry.
In particular
the couplings \coupl\ should obey the conditions
described in
the previous section for the couplings of $N=1, D=4$
supergravity.
There are two ways
to describe this depending on
the way we choose coordinates on the
moduli space which we shall now describe below.

\section{$G_2$ Moduli Space: A K\"ahler Approach}

The sigma model metric $ds^2$ in \coupl\ is required by
$N=1, D=4$
supersymmetry to be K\"ahler. In addition the action of the
group $\bR^{b_3}$ on $p$ leaves the
sigma model metric $ds^2$ invariant
and also commutes with the
supersymmetry transformations of the
scalars. This is because, as for the D=4 effective
action, the supersymmetry transformations
depend only on the spacetime
derivatives of the fields $p$. This
implies that $\bR^{b_3}$
acts with holomorphic isometries on the
sigma model target space $M$
which can be identified with the tangent
space $T{\cal M}$ of the
moduli space $\cM$ of $G_2$ structures.
The typical fibre of $T{\cal M}$
is $H^3(N, \bR)$.

To continue the investigation of the
moduli space geometry, it is convenient
to choose coordinates
on the moduli space $\cM$ so that the parallel form is
$$
\phi= s^i \phi_i\ .
\eqn\parfrm
$$
In fact, the basis $\phi_i$ of harmonic 3-form
with respect to the $G_2$
metric depends on the choice of $G_2$ structure
 and so on the coordinates $s$.
We take  the origin of the coordinate system to be
$s^i=s^i_o\not=0$.
However this dependence does not contribute
 in the couplings \coupl\ of four-dimensional
theory because, as we have mentioned in the
 previous section, they do not
depend on the choice of harmonic representatives for
$\phi_i$.
Next one can introduce holomorphic coordinates
$z^i=-{1\over2}(s^i+ip^i)$ on $T{\cal M}$
such that $\bR^{b_3}$ acts on $z^i$
with shifts along the imaginary directions.
In these coordinates, the sigma model metric on $T\cM$ is
$$
ds^2=\gamma_{i\bar j}dz^i d\bar z^{\bar j}=
\partial_i \partial_{\bar j} K({\rm Re} z) dz^i d\bar z^j\ ,
\eqn\smm
$$
and the K\"ahler form is
$$
\Omega= i dz^i\wedge  d\bar z^{\bar j}\partial_i
\partial_{\bar j} K({\rm Re} z)\ .
\eqn\kfrm
$$
Comparing the metric \smm\ with that in \coupl, we
find that
$$
4 k_{ij}=4 m_{ij}=\partial_i \partial_j K
={1\over \int_N d^7y {\sqrt G}}\,\int_N\,
{\sqrt{ G}}\, d^7y\, (\phi_i, \phi_j)\ ,
\eqn\sccpl
$$
where $K$ is the K\"ahler potential.

We turn now to investigate the couplings of the
 vector multiplets. Using
 the holomorphicity of $h_{ab}$ required by
supersymmetry and the expression
given in \coupl, we find that
$$
h_{ab}= z^i C_{iab}\ .
\eqn\gccpl
$$
The coupling $h_{ab}$ can be though as a holomorphic
section of a
bundle with fibre $S^2 H^2(N, \bR)\otimes \bC$ over
the sigma model
manifold $T{\cal M}$.

It remains now to find the K\"ahler potential of the metric
on the sigma model manifold $T\cM$.
We shall show that the K\"ahler potential
\foot{A similar expression for the K\"ahler
 potential, but with different
normalization factor,  was given in [\hm].} is
$$
K=-{3\over7}\log\big[\int_N \phi\wedge *\phi]=
-{3\over7}\log[\int_N\, d^7y\, {\sqrt G}\big]\ .
\eqn\kpot
$$
For this, we shall use the relation shown by Hitchin
 in [\hitchin] that
$$
\hat K=\int_N \phi\wedge {}^*\phi=
\int_N [{\rm det} B]^{{1\over 9}}\ ,
\eqn\hitrel
$$
where $[{\rm det} B]^{{1\over 9}}$ is a
top-form constructed taking the determinant of
$$
B_{IJ}=-{1\over 6\, 4!} \phi_{I K_1K_2}  \phi_{J K_3K_4}
\phi_{K_5 K_6 K_7}\,\, dy^{K_1}\wedge\dots \wedge dy^{K_7} \ .
\eqn\hitmat
$$
It is clear from this that $\hat K$ is
homogeneous of degree $7/3$ in the
$s$ coordinates\foot{In [\hitchin], the
metric on the moduli space $\cM$
is taken to be the Hessian of $\hat K$
which differs from the metric
that arises in the compactifications we are investigating.
The metric
associated with the Hessian of $\hat K$ has Lorentzian
signature.}
. The metric on the $G_2$ holonomy manifold is given by
$ G_{IJ}= ({\rm det} B)^{-{1\over9}} B_{IJ}$.
Using this one can show that
$$
{\partial\over \partial s^i}\int_N \phi\wedge {}^*\phi=
{7\over3}\int_N \phi_i\wedge  {}^*\phi\ .
\eqn\scal
$$
To proceed we remark that the $\Lambda^3(\bR^7)$
representation of $SO(7)$ of dimension
${\bf 35}$ can be decomposed
under the action of $G_2$ as ${\bf 1}+ {\bf 7}+ {\bf 27}$.
The first is the direction along the $G_2$ invariant
form $\phi$. The representation
${\bf 7}$ is associated with the vector arising from
the inner product of a 3-forms  with
$*\phi$. If the three-form is harmonic
 with respect to the $G_2$ metric,
as it is the case here, then this part
vanishes due to a standard argument
about harmonic one-forms on irreducible  Ricci-flat spaces.
Therefore  $\phi_i$ can be written as
$\phi_i=\pi_1(\phi_i)+\pi_{27}(\phi_i)$,
 where $\pi_1$ and $\pi_{27}$ are the
obvious projections.
In addition it is known [\joyce, \hitchin] that
$$
{\partial\over \partial s^i}{}^*\phi=
{4\over 3} {}^* \pi_1(\phi_i)- {}^*\pi_{27}(\phi_i)\ .
\eqn\scala
$$
 Using \scal\ and \scala\ it is straightforward to
verify \kpot.
In conclusion, we find
$$
k_{ij}=m_{ij}=- {3\over 28} \partial_i \partial_j
\log \int_N \phi\wedge *\phi\ .
\eqn\sccplb
$$
The numerical normalization factors that
appear in the expressions for the
moduli metric and K\"ahler potential are
important in the investigation of the various
brane solitons that arise in these compactifications.

We remark that the components $k_{ij}$ and $m_{ij}$
of the sigma model metric are homogeneous of degree $-2$
under the scaling of the $s$ coordinates of the
moduli space $\cM$.
This follows in a straightforward manner from the homogeneity
properties of the volume of $G_2$ manifolds that
 we have explained
above. We find that \smm\ is
invariant under scaling $s^i\rightarrow \ell s^i$ and
 $p^i\rightarrow \ell p^i$,
where  $\ell\in \bR-\{0\}$. So this scaling transformation
is an isometry. The isometry group of the metric
\smm\ on $T\cM$
is the semi-direct product of $\bR-\{0\}$
with $\bR^{b_3}$ the
group of translations along the $p^i$ coordinates.

\section{$G_2$ Moduli Space: A symplectic Approach}

An alternative way to describe the geometry of the moduli
space is to use symplectic geometry. For this we consider
the cotangent bundle $T^*{\cal M}$ of the moduli
space $\cM$, a typical fibre of which can be
identified with $H^4(N, \bR)$. There is a symplectic
pairing between
$H^3(N, \bR)$ and $H^4(N, \bR)$ given by Poincar\'e duality.
Of course $M=T{\cal M}$ is isomorphic to $T^*{\cal M}$.
Choose now coordinates on $T^*{\cal M}$ such that the
K\"ahler form is
$$
\Omega=  du^i\wedge dr_i\ .
\eqn\symplkf
$$
Next we write the  metric on $T^*\cM$ as
$$
ds^2=n_{ij}(u) du^i du^j+ q^{ij}(s) dr_i dr_j\ .
\eqn\extra
$$
 Given the symplectic
form and the metric, one can introduce  the (almost)
 complex structure
$$
\eqalign{
I^{s^i}{}_{r_j}&=n^{u^iu^k} \Omega_{u^kr_j}=n^{ij}
\cr
I^{r_i}{}_{u^j}&=q^{r_ir_k}\Omega_{r_ku^j}=-q_{ij}\ ,}
\eqn\cpxstr
$$
where $n^{ij}$ and $q_{ij}$ are the
inverse matrices of $n_{ij}$ and
$q^{ij}$, respectively.
Requiring that $I^2=-1$, we find that
$$
n_{ij}=q_{ij}\ .
\eqn\identaa
$$
It remains to investigate the integrability of
the above complex structure.
For this define the (1,0) forms
$$
\eqalign{
e_i&= idr_i-n_{ij} du^j
\cr
\hat e^i&=idu^i+ n^{ij} dr_j\ .}
\eqn\integra
$$
Observe that $ e_i= i n_{ij} \hat e^j$.
Requiring that $de_i$ and $d\hat e^i$ do not
contain a (0,2) part, we find that
$$
\partial_{[i} n_{j]k}=0
\eqn\intergrcnd
$$
which in turn implies that
$$
n_{ij}=\partial_{i} \partial_j \tilde K(u)
\eqn\intcndb
$$
for some function $\tilde K(u)$.
A set of complex coordinates with respect to the
above complex structure is
$$
\tilde z_i= -\partial_i \tilde K+ i r_i\ .
\eqn\cpxcoord
$$
To make connection with the K\"ahler approach to the geometry
of the moduli space, define the coordinates
$$
v_i=-\partial_i\tilde K\ , \qquad r_i=r_i\ .
\eqn\ccc
$$
Then the sigma model metric becomes
$$
ds^2=n^{ij} dv_i dv_j+ n^{ij} dr_i dr_j\ .
\eqn\sigmmm
$$
Next observe from \ccc\ that $du^i=-n^{ij} dv_j$. Taking the
exterior derivative of this equation, we find that
$$
{\partial\over \partial v_i} n^{jk}-
{\partial\over \partial v_j} n^{ik}=0
\eqn\extder
$$
which in turn implies that $n^{ij}$ can be expressed as two
$v$-derivatives on a scalar.
 Setting
$v_i=s^i$,  $r_i=p^i$ and $n^{ij}=k_{ij}$,  we establish
the relation
between the K\"ahler and symplectic approaches to the
geometry of $T\cM$.
The K\"ahler geometry  on $T\cM$ will be used in the
sections below
 to construct
solutions for the $N=1$, $D=4$ supergravity that arise from
compactifications of M-theory on holonomy
$G_2$ manifolds and are associated
with M-branes wrapped on cycles in the compact space.

\section{Potentials}

 As we have seen, potentials do not arise
in the four-dimensional effective
theory associated with the (direct) compactification
 of $N=1, D=11$ supergravity on $G_2$ holonomy manifolds.
 However several mechanisms
 have been proposed for generating a potential.
 One such mechanism involves compactifications in
  the presence of a non-vanishing
 4-form field strength $F_4^0$ along the directions of the
compact manifold. This is a Scherk-Schwarz type
 of mechanism which has been recently adapted
in the context of $G_2$
compactifications of string theory in [\gukov, \raf, \spen].
 It turns out though that the presence of non-vanishing
  $F_4^0$
in the context of $G_2$ compactifications of M-theory
is not consistent with the compactness of the
 internal manifold [\becker].
Nevertheless, adapting the formalism proposed in
[\gukov, \raf, \spen] we find that
the superpotential $f$ associated with such $F^0$ is
$$
{\rm Re} f=\int_N \phi\wedge F^0\ .
\eqn\poteta
$$
The imaginary part of $f$ is determined by holomorphicity.

 Other mechanisms of generating a potential in the
 low energy effective action in four dimensions involve
instanton
  effects which arise from wrapping
 M2-branes on associative 3-cycles of the $G_2$ manifold.
 Such cycles exists in some $G_2$ holonomy
manifolds and the associated
 instantons induce a scalar  potential.
 In particular it has been found in [\hm] that a
probe homology 3-sphere instanton
 M2-brane wrapping the cycle $C$ generates the superpotential
 $$
 f(z)=\m e^{k_i z^i}\ ,
\eqn\instpot
 $$
 where $\m$ is a constant and $k_i\sim \int_C \phi_i$.
 Such cycles exist in special holonomy $G_2$ manifolds
but they may not exist for generic ones (see next section).
In the case that such a superpotential \instpot\  appears
it easy to see using the results of section
three that the only supersymmetric vacuum
is at $|s|\rightarrow \infty$. The theory may
have other vacua but they are not supersymmetric.

 Because of the above mentioned difficulties
for generating a potential
 for generic $G_2$ compactifications of
M-theory, we shall mostly  focus
 in the investigation of the solutions
of the four-dimensional
 action with couplings described in
section three and without a potential.
 However from the perspective of the
general $N=1$, $D=4$ supergravity,
 one can do a more general analysis which
we shall present in an appendix.

\chapter{Cycles and Wrapped Branes}

\section{Calibrations and Supersymmetric Cycles}

On $G_2$ holonomy manifolds there are
 two calibrations that are associated
with supersymmetric cycles. One is of
degree three (associative)
calibration and the other of degree
four (coassociative) calibration
associated with the parallel three- and
four-forms on these manifolds [\hl].
We shall refer to them as supersymmetric calibrations.
There may be other calibrations on $G_2$-manifolds
but they will not be supersymmetric.
To see this, the supersymmetry condition which
is deduced from $\kappa$-symmetry
is [\strombeck,\kallosh]
$$
\Gamma\eta=\eta\ ,
\eqn\scstr
$$
where $\Gamma$ is the $\kappa$-symmetry
projector and $\eta$ is
the $N=1,D=11$ supersymmetry parameter which
should be parallel with respect
to the Levi-Civita connection of the compact
$G_2$ holonomy manifold.
Since from such a parallel spinor $\eta$, one
can construct the parallel three- and
four-forms, only calibrations associated
 to these forms  are supersymmetric.

Even though supersymmetric calibration forms exist
on $G_2$ manifolds, it is not apparent
that there  always exist
(calibrated) supersymmetric  representatives
of the homology
3- and 4- classes of the  $G_2$ manifold,
respectively. It is known that if
supersymmetric (associative) 3-cycles exist,
  they do not have moduli [\macl].
Thus such 3-cycles are isolated. In fact  it has been
conjectured that they do not exist
for generic $G_2$ manifolds, though
one expects to find them in special cases[\joyceb].
Supersymmetric, coassociative,  4-cycles, $X$,
 have moduli in $G_2$ manifolds
with dimension\foot{We use conventions similar to [\joyceb].}
  $b^2_+(X)$  [\macl]. So one expects to have locally
 smooth moduli space for  coassociative  calibrations.

There are several homology cycles on
$G_2$ manifolds on which one can wrap
M-theory branes. We shall be mainly
concerned with two-cycles, three-cycles,
four-cycles and five-cycles. As we have seen,
two- and five- cycles cannot be supersymmetric.
This however does not mean that none
of them is calibrated, with respect
to a non-supersymmetric calibration,
 or there is no minimal submanifold
in the homology class of these cycles.
Three- and four-cycles can be supersymmetric,
but as it has been mentioned above this does not
necessarily imply that every three- and four-cycle has
a supersymmetric calibrated submanifold  representing
its homology class.

\section{ Wrapping M-branes on Homology Cycles}

The brane solitons in four dimensions that one
expects to find by wrapping
M-branes on homology cycles of
 $G_2$ holonomy manifold $N$  which are represented by
a minimal submanifold are as follows:

\item{(i)} Wrapping  M2-branes on two-cycles leads
 to  non-supersymmetric
0-branes in four dimensions.

\item{(ii)}Wrapping M5-brane on two-cycles,
three-cycles, four-cycles and five-cycles leads to
non-supersymmetric spacetime filling 3-branes,
 supersymmetric 2-branes,
supersymmetric 1-branes and non-supersymmetric 0-branes,
respectively.

All brane configurations above that
arise from wrapping M-branes to four-dimensions
should be described by  solutions of
the effective $N=1, D=4$ effective supergravity
theory of this compactification.
Since spacetime filling supersymmetric or
non-supersymmetric 3-branes are characterized
by $3+1$-dimensional Poincar\'e invariance,
the associated supergravity
solutions are those of flat Minkowski spacetime
with vanishing gauge potentials
and constant scalars. Such solutions  are
of course the (supersymmetric) vacua of the theory.

Some non-supersymmetric 0-brane solutions
can be identified with the black hole solutions
of the supergravity theory. Typically
the electrically charged
black holes are associated with wrapped M2-branes, and
magnetically charged ones with wrapped M5-branes.
The 2-cycles and 5-cycles in the $G_2$ holonomy manifold
are Poincar\'e dual to each other.
It is well known that the electrically and magnetically
charged black holes in four dimensions  are dual to each
other
via electromagnetic duality. So one can view the
electro-magnetic duality
in four-dimensions as consequence of the Poincar\'e
duality on $G_2$ manifolds.

The 1-brane configurations can be identified with strings.
As we shall see the solutions are in fact similar to those of
cosmic strings [\cg, \vy].
The 2-brane  solitons  are the domain wall
solutions of $N=1, D=4$ supergravity theory.

\chapter{Strings}

\section{Coassociative Cycles and M5-branes}

As we have mentioned the string solutions
of $N=1$, $D=4$ supergravity
can be thought off as M5-branes wrapped
on coassociative cycles of the $G_2$
manifold. It is known that the supersymmetry
conditions [\gwggp, \west] associated
with such a cycle in  the directions $123457$ are
$$
\eqalign{
\Gamma_{1346} \epsilon&=\epsilon
\cr
\Gamma_{2356} \epsilon&=\epsilon
\cr
\Gamma_{4567} \epsilon&=\epsilon\ .}
\eqn\gtp
$$
Now suppose that we place a M5-brane extended in the
directions
$081346$ associated with the projection
$$
\Gamma_{081346}\epsilon=\epsilon\ .
\eqn\proja
$$
The string directions are taken to be $08$.
It is easy to see that the above projectors
lead to a configuration
that preserves two supersymmetries, ie it preserves $1/2$
of supersymmetry of $N=1$, $D=4$ theory.

There are two simple cases of coassociative
cycles to consider. One is
that of coassociative cycles with the topology
of the torus $T^4$ and
the other is of coassociative cycles with the topology
of a $K_3$ surface. In both
cases the dimension of the moduli space is three.  So
one expects to find string solutions of $N=1, D=4$
supergravity
 associated with the wrapping of
M5-branes on these coassociative  cycles with
 topology $T^4$ and $K_3$.
 The tension of corresponding  strings
will be  equal to the tension of the M5-brane
times the volume of the
coassociative  cycles.

\section{$G_2$ Strings}

To investigate the string solutions to the supergravity field
equations it is convenient to use the K\"ahler parameterisation
of the moduli space of $G_2$ structures. The $G_2$ strings are
a special case of the cosmic strings for which the sigma model
manifold is the space $T\cM$.
The solution is
$$
\eqalign{
ds^2&= ds^2(\bR^{1,1})+ e^{-K} dw d\bar w
\cr
z^i&=z^i(w)
\cr
A^a&=0\ ,}
\eqn\strnganz
$$
where $w$ is a complex coordinate of spacetime, and
the  K\"ahler
potential $K$ and the complex coordinates $z^i$ are given
 in section three.
The $G_2$ strings above do not have finite tension because
$T\cM$ is not compact. However it is known that the
 fibre directions
of $T\cM$ can be compactified to a torus by
dividing with $H^1( T^{b_3}, \bZ)$
which is thought of as a group of large gauge
 transformations.
This leaves the base
 $\cM$ of $T\cM$ which is an open set in $H^3(N, \bR)$.
 It is expected that certain points of $\cM$
 should be identified
by large diffeomorphisms of the compact $G_2$ holonomy
  manifold $N$. However to our knowledge it is
not known how such large diffeomorphisms act on $\cM$ and
 whether
they are sufficient to allow for string
 solutions with finite  tension
 adapting a similar construction in [\vy].
The identification of the $G_2$ strings
 with wrapped M5-branes
on coassociative  cycles provides some
indirect evidence that $G_2$ strings
with finite tension exist in the
 case when coassociative cycles
are present. We note that there
 are supersymmetric string solutions
even in the presence of a Fayet-Iliopoulos term [\jggp].

\chapter{$G_2$ Domain Walls}

The supersymmetry projections of domain walls that arise from
a M2-brane in the directions
$089$ in the
background of the $G_2$ manifold in the
directions $1234567$, are those of
\gtp\ and
$$
\Gamma_{089}\epsilon=\epsilon\ .
\eqn\projb
$$
So the supersymmetry preserved is $1/16$ of M-theory, ie
$1/2$ of that of $N=1, D=4$ supergravity.
 The domain wall is in the directions $089$.

It is straightforward to write
 the supergravity solution of a brane that is located
in a Ricci-flat manifold.
For the case of interest  here,
the transverse space of the  M2-brane is
$\bR\times N$, where $N$ is the holonomy $G_2$
compact space. The solution is
$$
\eqalign{
ds^2&=h^{-{2\over 3}} ds^2(\bR^{1,2})+
h^{1\over3} (ds^2_{(7)}+dy^2)
\cr
F_4&={1\over2} dvol(\bR^{1,2})\wedge dh^{-1}\ ,}
\eqn\lmtwo
$$
where $h$ is harmonic in $\bR\times N$. Additional
fluxes $F^0$
can be added in the solution along $\bR\times N$.
However in this case
$h$ is not harmonic but rather obeys the equation
$\Delta h=|F^0|^2$. For $F^0=0$,  $h$ can be chosen
to be harmonic in $\bR$, ie $h$ is
piece-wise linear function of the
coordinate $y\in \bR$, see [\wipo, \ptb].
This solution has electric fluxes and it
cannot be described from the four-dimensional
perspective using the linearized compactification
ansatz of section three.
We shall not pursue this point further here.

Alternatively,  domain walls can arise by wrapping M5-branes
on associative 3-cycles of the compact space.
If the $G_2$ manifold is
in the directions $1234567$ and the
M5-brane is in the directions
$012389$, then the projections are as in \gtp\ and in addition
$$
\Gamma_{012389}\epsilon=\epsilon\ .
\eqn\projc
$$
These lead again to a configuration
 preserving $1/16$ of
M-theory supersymmetry, ie $1/2$ of
 supersymmetry of
$N=1, D=4$ supergravity. From the
 perspective of $N=1, D=4$
supergravity, these domain walls
 may arise from
a superpotential generated by a
M2-brane instanton wrapping
the associative cycle.  The
investigation of
the Killing spinor equations of $G_2$
domain walls that preserve $1/2$
of supersymmetry  will
be given in an appendix.

\chapter{Black Holes}

\subsection{Black Holes as Wrapped M2-branes}

As we have mentioned the electrically charged black holes
of $N=1$, $D=4$ supergravity action that arise  from
compactifying M-theory on holonomy $G_2$ manifolds can
be viewed as wrapped M2-branes on the two-cycles of the
compact space $N$. It is known that there are minimal
sphere representatives of every homotopy class $\pi_2(N)$
(see for example chapter VI  [\schoen]).
The mass and the charges
of such black holes are given by
$M=T_2 Vol(C)$ and
$Q_a=T_2 \omega_a[C]$, respectively, where $T_2$
is the M2-brane tension and $C$ is the two-cycle.

Since two cycles in holonomy $G_2$ manifolds are not
supersymmetric, it is not expected
to find  relation
between the mass and the charges of the black holes. This
is despite the fact that the mass and the charge per-unit
 volume of the
associated M2-brane are equal.
Although it is not apparent that there are non-supersymmetric
degree two calibrations on holonomy $G_2$ manifolds, suppose
that there is one associated with the two-form $\lambda$.
Then $\lambda=\xi^a \omega_a$,  for some constants $\xi^a$,
and
$M=T_2 Vol(\Sigma)=T_2 \int_{C} \phi=T_2 \xi^a Q_a$
which can be interpreted as an extremality relation for the
black hole. As we shall see there are extreme solutions
 $N=1, D=4$ supergravity associated with $G_2$
compactifications  which however exhibit a naked singularity.

\section{Electric  $G_2$ Black Holes}

\subsection{Ansatz and Field Equations}

In order to find  black hole solutions of $N=1, D=4$
supergravity
associated with $G_2$ compactifications of M-theory, we
 consider the ansatz
$$
\eqalign{
ds^2&=-A^2(r) dt^2+ B^2(r) (dr^2+ r^2 ds^2(S^2))
\cr
A^a&= dt C^a(r)
\cr
s^i&= s^i(r)
\cr
p^i&=p^i(r)\ .}
\eqn\bhans
$$
We recall from section 3 that $\gamma_{i\bar j} dz^i
d\bar z^j$$= k_{ij} (ds^i ds^j+dp^i dp^j)$
with $k_{ij}=$$\pd{i}\pd{j} \Phi(s)$, where
$\Phi = -{3 \over 28} \log (\Theta (s))$,  $\Theta$ is
the volume
of the compact space and $\pd{i} = {\partial \over \partial s^i}$
; so $\pd{i} \Phi$ is
homogeneous of degree $-1$ in $s$.  Furthermore we shall
take the scalar potential
of $N=1, D=4$ supergravity to vanish $V \equiv 0$.
It is straightforward to observe from the
Killing spinor equations that all  electrically charged
solutions cannot be supersymmetric.

Substituting the ansatz \bhans\ into the Maxwell equations
 we find
$$
B A^{-1} r^2 \rre h_{a b} \pd{r} C^b = H_a\ ,
$$
where $H_a$ are constants.
To obtain the scalar equations we vary $s^\ell$ and $p^\ell$.
Recalling  that $\iim h_{ab} = -{1 \over 2} C_{iab}p^i$ and
 $\rre h_{ab} =-{1 \over 2}  C_{iab} s^i$,
we find that the field equations for $s^i$ and $p^i$ are
$$
\eqalign{
\sqrt{|g|} \big( {1 \over 8} C_{\ell ab} F{}^a{}_{MN} F^{bMN}
- \pd{\ell}
k_{ij} \big( \pd{M} s^i \pu{M} s^j + \pd{M} p^i \pu{M}
p^j \big) \big)
\cr
+2 \pd{M} \big( \sqrt{|g|} k_{\ell j} \pu{M} s^j \big) =0}
\eqn\bha
$$
and
$$
-{1 \over 8} \sqrt{|g|} C_{\ell ab}  F{}^a{}_{MN}
{}^\star F^{bMN}
+2 \pd{M}\big( \sqrt{|g|} k_{\ell j} \pu{M} p^j \big) =0\ ,
\eqn\bhb
$$
respectively.

It is convenient to define
$$
\psi = AB \qquad \qquad N=rB\ .
\eqn\psni
$$
Then the Maxwell equations may be rewritten as
$$
N^2 \psi^{-1} \rre h_{ab} \pd{r} C^b = H_a \ .
\eqn\gags
$$

On substituting the black hole ansatz into
the scalar equations, and eliminating $\pd{r} C^a$
using \gags,  one obtains
$$
\eqalign{
 -\psi r^2 \pd{\ell} k_{ij} \big(  \pd{r} s^i \pd{r}s^j
+ \pd{r} p^i \pd{r} p^j \big)
\cr -{1 \over 4} \psi N^{-2} C_{\ell ab}
\rre h^{ac} H_c \rre h^{bd} H_d
+2 \pd{r} (\psi r^2 k_{\ell j} \pd{r} s^j \big) =0
\cr
\pd{r} \big( \psi r^2 k_{\ell j} \pd{r} p^j \big) =0 \ .}
\eqn\bhc
$$

Lastly we consider the Einstein equations. We adopt
the notation ${\dot{B}} = {d B \over dr}$,
${\dot{A}} = {d A \over dr}$, $\ddot{B}  =
{d^2 B \over dr^2}$, $\ddot{A} = {d^2 A \over dr^2}$.
The non-vanishing components of the Einstein tensor are
given by
$$
\eqalign{
G_{00} & = -{A^2 \over rB^4} \big( 4B \dB -r \dB^2 +2rB \ddB
\big)
\cr
G_{rr} & = {1 \over rAB^2} \big(2r\dA \dB B+2AB \dB
+rA\dB^2 +2B^2 \dA \big)
\cr
G_{\theta \theta} & = {r \over AB^2} \big(B^2 \dA +BA\dB
+ABr \ddB +rB^2 \ddA -rA \dB^2 \big)
\cr
G_{\phi \phi} & = {r \sin^2 \theta \over AB^2}
\big(B^2 \dA +BA\dB +ABr \ddB +rB^2 \ddA -rA \dB^2 \big) \ .}
\eqn\bhd
$$
Eliminating  $C^a$ from the above equations
using \gags\ and utilizing the definitions \psni,
we find that  the independent
Einstein equations can be expressed as
$$
{d^2 \over dr^2} (r^{3 \over 2} \psi) =
{3 \over 4} r^{-{1 \over 2}} \psi
\eqn\eina
$$
$$
{\psi \over N} {d \over dr} ({{\dot N} \over \psi})
= -k_{ij} \big( \pd{r} s^i \pd{r} s^j
+\pd{r} p^i \pd{r} p^j \big)
\eqn\einb
$$
$$
r^{1 \over 2} N^3 \psi^{-1} {d^2 \over dr^2}
(N^{-1} r^{3 \over 2}
\psi) +N \big[ {5 \over 4}N-r {\dot N}-r^2{\ddot N} \big]
- \rre h^{ab} H_a H_b =0\ .
\eqn\einc
$$
A useful identity implied by the scalar equations (or
the Einstein equations) is
$$
-{1 \over 2} \psi^2 r^2 N^{-2} \pd{r} \big(
\rre h^{ab}H_a H_b \big)
+ \pd{r} \big(\psi^2 r^4 k_{ij} \big[ \pd{r} s^i \pd{r} s^j
+  \pd{r} p^i \pd{r} p^j \big] \big)=0\ .
\eqn\uid
$$

Using the Einstein equation \eina, we determine  $\psi$ as
$$
\psi = \b + {\a \over r^2}
\eqn\nnn
$$
for constants $\a$, $\b$. Then  \einc\ may be simplified to
$$
2 \psi -2 {d \over dr} \big( \psi r^2 N^{-1} {\dot N} \big)
 = \psi N^{-2} \rre h^{ab} H_a H_b\ .
\eqn\crumbs
$$
In addition, from the $p^\ell$ scalar equation, $p^\ell$
 must satisfy
$$
\pd{r} p^\ell = \psi^{-1} r^{-2} k^{\ell j} \theta_j
\eqn\bhe
$$
for some  real constants $\theta_j$, and
 here $k^{\ell j}$ is the inverse of the Hessian of $\Phi$.

A solution to the above system of equations is the
Schwarzschild black hole. In this case
the Maxwell gauge potential vanishes and the
scalars are constant.
This black hole cannot be thought of as a M2-brane
wrapped on a homology 2-cycle
of the compact $G_2$ holonomy manifold because
it does not carry electric charges.
 Generically, to find  electrically
charged black holes, one has to take some of the scalars
to be non constant\foot{In special
cases this depends on the properties of the intersection
 numbers $C_{iab}$ but we shall
not pursue this further here.}.
So we take $\pd{r} s^i \neq 0$. On eliminating $p^i$
from the scalar
equation for $s^\ell$ and making use of the fact that
$k_{ij}$ is the Hessian of $\Phi$, one obtains
$$
\eqalign{
-{1 \over 4} \psi N^{-2} C_{\ell ab}
\rre h^{ac} \rre h^{bd} H_c H_d
+ \psi r^2 \pd{\ell} k_{ij} \pd{r} s^i \pd{r} s^j
\cr
 + \psi^{-1} r^{-2} \pd{\ell} k^{ij} \theta_i \theta_j
+2 k_{\ell j} \pd{r} \big( \psi r^2 \pd{r} s^j \big) =0\ .}
\eqn\simpsc
$$
To solve the field equations, we shall make use
 of the homogeneity properties  of $\pd{i} \Phi$.
In particular, contracting \simpsc\ with $s^\ell$
and setting $\theta_i=0$,
one finds the identity
$$
{1 \over 2}\psi N^{-2} \rre h^{ab} H_a H_b -
2 \pd{r} \big( \psi r^2 \pd{r} \Phi \big)=0 \ .
\eqn\bhh
$$
From this it follows that in order to have charged solutions,
we shall require
$\pd{r} (\psi r^2 \pd{r} \Phi) \neq 0$.
Substituting \bhh\ into \crumbs\ with $\psi=1-{\m^2 \over r^2}$
for $\m \neq 0$, one obtains
$$
N = \k r \psi e^{-2 \Phi} \big( {r + \m \over r-\m}
\big)^{\d}
\eqn\bhg
$$
for constants $\k$ and $\d$.
In the special case when $\psi =1$ one obtains
$$
N = \k r e^{-2 \Phi} e^{{\d \over r}} \ .
\eqn\bbhg
$$
The {\sl  only remaining independent equations}
are  the Einstein equation \einb\ together with \simpsc\ .
We have been unable to
find general solutions to the field equations,
and our general analysis ends here.

\subsection{Dilatonic $G_2$ Electric Black Holes}

To find an explicit solution to the equation
\einb\ and \simpsc, we shall now consider
the special case where
$$
p^i=0 \qquad \qquad s^i=s(r) c^i
\eqn\bhh
$$
for some constants $c^i$.
Then the homogeneity of $\pd{i} \Phi$ implies that
$$
\eqalign{
\pd{i} \Phi &= \l_i s^{-1}
\cr
k_{ij}& = \pd{i} \pd{j} \Phi = \l_{ij} s^{-2}
\cr
\pd{\ell} k_{ij} &= \pd{i} \pd{j} \pd{\ell} \Phi =
\l_{ij\ell} s^{-3}}
\eqn\scala
$$
where $\l_i$, $\l_{ij}$ and $\l_{ij\ell}$ are
constants related by
$$
\eqalign{
c^i\l_i&=-{1\over4}
\cr
c^i \l_{ij}& =- \l_j
\cr
c^i \l_{ij\ell}& = -2 \l_{j \ell}\ .}
\eqn\scalab
$$
It is also convenient to define $\cH_{ab} =
-{1 \over 2}c^\ell C_{\ell ab}$ so that
$\rre h_{ab} = s \cH_{ab}$.  Then the $s^\ell$ scalar
 equation \simpsc\  implies that
$$
-{1 \over 2} C_{\ell ab} \cH^{ac} \cH^{bd} H_c H_d =
 \l \l_\ell
\eqn\constid
$$
for some constant $\l$, and
$$
{\l \over 2} \psi N^{-2} s^{-2} +2 \psi r^2 s^{-3} (\pd{r}s)^2
-2 s^{-2} \pd{r} \big(\psi r^2 \pd{r}s \big)=0 \ .
\eqn\bhi
$$
Note that on contracting \constid\ with $c^\ell$ and
using the homogeneity properties
of $\Theta$,  one obtains
$$
-{\l\over 4}  = \cH^{ab} H_a H_b \ .
\eqn\bhj
$$
 Then using the above identities,
it follows that for $\pd{r} s \neq 0$ the scalar
 equations are implied by the Einstein
equations. It therefore suffices to solve
$$
\eqalign{
{\psi \over N} {d \over dr} ({{\dot N} \over \psi})& =
-{1\over4}  s^{-2} (\pd{r} s)^2
\cr
2 \psi -2 {d \over dr} \big( \psi r^2 N^{-1} {\dot N} \big)& =
-{1\over4} \psi N^{-2}
s^{-1} \l  \ .}
\eqn\simpac
$$
To find solutions to these equations which are
asymptotically Minkowski, we shall set $\psi=1-
{\m^2 \over r^2}$,
 see \nnn.
The simplest way to solve \simpac\ is to eliminate $s(r)$
from the first
equation by making use of the second equation. This equation
 for $N(r)$ may
be simplified further by defining $H(r)$ according to
$$
{\dot N \over N} = r^{-2} \psi^{-1} \big( H +r +
{\m^2 \over r} \big) \ .
\eqn\bhk
$$
Then $H$ must satisfy
$$
\eqalign{
(r^2-\m^2)^2 {\ddot H}^2 +4r (r^2-\m^2){\dot H}
{\ddot H}+ 4(r^2- \m^2) H {\dot H} {\ddot H}
\cr
 + 4(r^2-\m^2) {\dot H}^3 +4 {\dot H}^2 \big[r^2+2rH
-4\m^2+2H^2 \big] =0\ .}
\eqn\sssmp
$$
We discard the solution $H = {\rm const}$ because it
 is inconsistent with \simpac\ for $\l \neq 0$.
To simplify this equation even further, define $X=H$
and define implicitly
$$
f(X) = ((r^2- \m^2) {\dot H})^{-1}\ .
\eqn\bhl
$$
Then
$$
{\ddot H} = -{1 \over (r^2-\m^2)^2} \big(2r f^{-1}
+f^{-3} f' \big)\ ,
\eqn\bhm
$$
where $f' = {d f \over dX}$. Substituting these expressions
into \sssmp\ , one obtains
$$
4 f^3 +8f^4 X^2-16 \m^2 f^4 + (f')^2 -4f^2X f' =0 \ .
\eqn\da
$$
On setting $f(X)= (L(X)-X^2)^{-1}$, this simplifies to give
$$
4L -16 \m^2 + (L')^2 =0\ .
\eqn\db
$$
There are two possible solutions to this equation,
$$
L(X) = 4 \m^2
\eqn\ssa
$$
or
$$
L(X) = 4 \m^2 -(X+\t)^2
\eqn\ssb
$$
for some constant $\t$. However, $L(X) = 4 \m^2$
leads to  $s$
constant and so  the Schwarzschild black hole. We
therefore take
$$
L(X) = 4 \m^2 -(X+\t)^2\ .
\eqn\dc
$$
Hence $H$ must satisfy
$$
(r^2 - \m^2) {\dot H} = 4 \m^2 -2H^2-2 \t H - \t^2\ .
\eqn\dd
$$
For $\m \neq 0$ there are three cases to consider,
according as to whether
$\t^2 = 8 \m^2$, $\t^2 > 8 \m^2$ or $\t^2 < 8 \m^2$.
We shall focus on
the cases $\t = \m =0$ and $\t^2 < 8 \m^2$.

For $\m=\t=0$, we find that
$$
\eqalign{
ds^2&= -(1+{2 \k^2 \over r})^{-1} dt^2 + (1+
{2 \k^2 \over r}) d {\bf{x}}^2
\cr
F^a&=-{16 \k^4 \over \l (r+2 \k^2)^2} \cH^{ab}H_b dr\wedge dt
\cr
s^i&= -{\l \over 16 \k^2 r}(2+ \k^{-2} r)  c^i}
\eqn\dfh
$$
for constant $\k>0$. This solution is
asymptotically Minkowski,
and the charges, mass and asymptotic values of the scalar are
$$
\eqalign{
Q^a&=s^{-1}_\infty  \cH^{ab} H_b
\cr
M&=\k^2
\cr
s_\infty&=-{\l \over 16\k^4}\ ,}
\eqn\cmss
$$
respectively.

The spacetime geometry of the  solution \dfh\
is that of two M-brane
configuration reduces to four dimensions on a 7-torus,
for example two orthogonally
intersecting M2-branes on a $0$-brane [\gpb].
The above solution exhibits a naked singularity at $r=0$.

For  $\t^2 < 8 \m^2$,  we find that
$$
H = -{1 \over 2} (\t + \sqrt{8 \m^2 - \t^2})+
{\sqrt{8 \m^2 - \t^2} \over 1-
\rho \big( {r+\m \over r-\m} \big)^{{ \sqrt{8 \m^2 -
\t^2} \over \m}}}
\eqn\dfe
$$
which in turn implies that the solution is
$$
\eqalign{
ds^2&= -r^2 \psi^2  N^{-2} dt^2 + r^{-2} N^2
\big[ dr^2 + r^2  ds^2 (S^2) \big]
\cr
F^a&= \psi N^{-2} s^{-1} \cH^{ab} H_b dr\wedge dt
\cr
s^i&=s(r) c^i}
\eqn\bhn
$$
where
$$
\eqalign{
N &= {\k \over r} (r^2 - \m^2) \big( {r+\m \over r-\m}
\big)^{{1 \over 4\m}(\t - \sqrt{8 \m^2 - \t^2})}
 \sqrt{-1+ (1+\k^{-2}) \big( {r+\m \over r-\m}
\big)^{{ \sqrt{8 \m^2 - \t^2} \over \m}}}
\cr
s& = -{\l \over 16 \rho \k^2 (\t^2 - 8 \m^2)}
\big[ \big( {r+\m \over r-\m} \big)^{-{1 \over 2 \m}
(\t + \sqrt{8 \m^2 - \t^2})}- (1+\k^{-2})
\big( {r+\m \over r-\m} \big)^{-{1 \over 2 \m}
(\t - \sqrt{8 \m^2 - \t^2})} \big]}
\eqn\dff
$$
for constant $\k>0$, and we have set $\rho = 1+\k^{-2}$ in
\dfe\ .

This solution is again  asymptotically Minkowski
as $r\rightarrow \infty$, and
the electric charges $Q^a$, the mass $M$ and asymptotic
value $s_\infty$
of the moduli scalar are
$$
\eqalign{
Q^a&=  s_\infty^{-1}  \cH^{ab} H_b
\cr
M&= {1 \over 2}(\t + (1+2 \k^2)\sqrt{8 \m^2 - \t^2})
\cr
s_\infty&={\l \over 16 \k^2 (1+\k^2)(\t^2-8 \m^2)}\ ,}
\eqn\mcsa
$$
respectively. To examine this spacetime geometry
 it is particularly
useful to consider the metric in the neighbourhood
 of $r=\m$. We
define $W=r - \m$. Then as $W \rightarrow 0^+$,
$$
N \rightarrow \rho W^{1- {1 \over 4\m}
(\t+\sqrt{8 \m^2 - \t^2})}
\eqn\bho
$$
where $\rho$ is determined by
$$\rho ={\k \over \m} \sqrt{1+\k^{-2}} (2\m)^{1+{1 \over 4 \m}
(\t + \sqrt{8\m^2-\t^2})} \ .
\eqn\bhp
$$
The leading order behaviour of the metric in the
neighbourhood of $W=0$ is given by
$$
ds^2 = -{1 \over 4} \rho^{-2} W^{{1 \over 2\m}
(\t + \sqrt{8 \m^2- \t^2})}dt^2
+ \m^{-2} \rho^2 W^{2 - {1 \over 2\m}(\t + \sqrt{8 \m^2-\t^2})}
 (dW^2+\m^2 ds^2 (S^2)) \ .
\eqn\asmet
$$
{}From this metric it is straightforward to show that
if there exist geodesics along which in-falling
particles may pass through $r=\m$ in a finite proper
time then it is
necessary to impose the constraint ${1 \over 4 \m}
(\t+ \sqrt{8 \m^2-\t})>0$.
This condition
is sufficient to ensure the positivity of the
mass given in \mcsa\ .
Furthermore, the Ricci scalar associated with \asmet\ is
 given by
$$
R = 2 \rho^{-2} \big[ \m^2 ( \d^2 -1)-W^2 \big] W^{2 \d -4}\ ,
\eqn\bhq
$$
where $\d = {1 \over 4 \m}(\t + \sqrt{8 \m^2 - \t})$.
In order for this to remain
bounded as $W \rightarrow 0$, one must impose the
condition $\d =1$, i.e. $\t = 2 \m$.
Hence, in order for there to be a horizon at $r=\m$
one requires $\t = 2 \m$.
For this special case, it is most convenient to change
 co-ordinates by setting
$$
v = {1 \over r}(r+\m)^2\ ,
\eqn\bhr
$$
so that the metric can be written as
$$
ds^2 = -L(v) F(v) dt^2 +L(v)^{-1} \big[ F(v)^{-1} dv^2 +
v^2 ds^2 (S^2) \big]
\eqn\bhr
$$
where
$$
F(v) = 1- {4 \m \over v} \ ,
\eqn\bhs
$$
and
$$
L(v) = \big( 1+ {4 \k^2 \m \over v} \big)^{-1}\ .
\eqn\bht
$$
This four-dimensional spacetime geometry has been
considered before in [\tseytlin]
where it was obtained via compactification of
 two intersecting
 non-extreme  M2-branes  on a 7-torus, where the two M2-branes have
the same electric charge. The  solution
\bhr\  has finite horizon area.

\section{Magnetic $G_2$ Black Holes}

To find magnetic black hole  solutions which
are associated with M5-branes wrapped on 5-cycles
of the compact space, we take
the ansatz for the metric and moduli scalars
as in the electric case \bhans\ but for the Maxwell
field we write
$$
F^a = \m^a \sin \theta d \theta \wedge d \phi
\eqn\bhma
$$
where  $\m^a$ are constants and $\phi, \theta$ are
the standard angular coordinates
on a two sphere.
We shall see that the electric and the magnetic
 solutions are related
as expected because  of electro-magnetic
duality in four-dimensions. So we shall
not elaborate in the description  of this case.

First it is straightforward to see that the
gauge field equations are satisfied provided that
$p^i = {\rm const}$. Taking this to be the case,
the scalar equations of $p^\ell$ are automatically
satisfied. Given this, the $s^\ell$ scalar equations
are given by
$$
{1 \over 4} \psi N^{-2} C_{\ell a b} \m^a \m^b +
\psi r^2 \pd{\ell} k_{ij} \pd{r} s^i \pd{r} s^j
+2 k_{\ell j} \pd{r} (\psi r^2 \pd{r} s^j) =0\ ,
\eqn\bhmb
$$
where $N, \psi$ are defined as in the electric case.
The Einstein field equations with vanishing scalar
potential imply
$$
\eqalign{
{d^2 \over dr^2} \big( r^{3 \over 2} \psi \big) =
{3 \over 4} r^{-{1 \over 2}} \psi
\cr
{\psi \over N} {d \over dr} \big( {{\dot N}
\over \psi} \big) = -k_{ij} \pd{r} s^i \pd{r} s^j
\cr
2 \psi - 2 {d \over dr} \big( \psi r^2 N^{-1}
{\dot N} \big) = \psi N^{-2} \rre h_{ab} \m^a \m^b\ .}
\eqn\bhmc
$$
The first two of these equations are identical
to the electric black hole case.
 In particular, the first equation implies
$\psi = \b +{\a \over r^2}$
for real constants $\a$, $\b$.

Next write  $s^i = s(r) c^i$ for some constants $c^i$. Using
\scala\ and \scalab\ and the definition of $\cH_{ab}$ as
in the electric case,  the $s^\ell$ scalar equation
 implies that
$$
-{1 \over 2}C_{\ell ab} \m^a \m^b = \l \l_\ell
\eqn\bhmd
$$
for some constant $\l$.  So it follows that
$\cH_{ab} \m^a \m^b = -(1/4)\l $.
Then just as for the electric case, by making use
of the identity
$$
-{1 \over 2} \psi^2 r^2 N^{-2} \pd{r} (\rre h_{ab} \m^a \m^b) +
 \pd{r} (\psi^2 r^4 k_{ij} \pd{r} s^i \pd{r} s^j) =0
\eqn\bhme
$$
obtained from the Einstein equations, one
may see that the scalar equations
are implied by the Einstein equations.
Hence, it suffices to solve the remaining Einstein equations
$$
\eqalign{
{\psi \over N} {d \over dr} ({{\dot N} \over \psi})&=
-{1\over4}  s^{-2} (\pd{r} s)^2
\cr
2 \psi -2 {d \over dr} \big( \psi r^2 N^{-1} {\dot N} \big)
&=-{1\over4} \psi N^{-2}
s \l \ .}
\eqn\simpacb
$$
 Clearly, these equations are equivalent to
\simpac\ under the transformation $s \rightarrow s^{-1}$.
Hence it follows that the
spacetime geometry of these magnetic black holes is
identical to the
electric black hole case, and $s_{\rm{magnetic}}
= s^{-1}{}_{\rm{electric}}$.
For example the M-theory interpretation of the
analogue of the solution \dfh\
is that of two M5-branes orthogonally intersecting
on a 3-brane [\gpb].

\vskip 1.0cm
\noindent{\bf Acknowledgments:} G.P. thanks ITP and
the Physics
Department of University of Stanford for hospitality.
G.P. is supported by a University Research
Fellowship from the Royal Society. J.G. is supported
by a EPSRC postdoctoral grant.
This work is partially supported by SPG grant PPA/G/S/1998/00613.

\APPENDIX{A}{A: Spinor Notation}
It is most convenient to present the
 supersymmetry transformations
in terms of a 4-component Majorana spinor $\e$ with
 real components.
We take  $\sigma^{{{M}}}=(\sigma^{{M}}{}_{\alpha\dot\beta})$
to be the Pauli matrices;
$$
\eqalign{
\s^{{0}} = \pmatrix{-1 \quad \ \  0 \cr \ 0 \quad -1}
\qquad \qquad
\s^{{1}} = \pmatrix{0 \quad \ \  1 \cr 1 \quad \ \ 0}
\cr
\s^{{2}} = \pmatrix{0 \quad -i \cr i \quad \ \ 0}
\qquad \qquad
\s^{{3}} = \pmatrix{1 \quad \ \ 0 \cr 0 \quad -1}}
\eqn\apa
$$

We set $\eta_{\underline{M} \underline{N}}=
 {\rm{diag}} (-1,1,1,1)$,
$\epsilon^{12}=\epsilon_{21}=1$, and to perform
the supersymmetry
calculations we define explicitly
$$
\eqalign{
\Gamma_{\ubx} = \pmatrix{ 0 \qquad \s^1 \cr \s^1 \qquad 0}
\qquad \qquad
\Gamma_{\uby} = \pmatrix{ -1 \qquad 0 \cr 0 \qquad 1}
\cr
\Gamma_{\underline{0}}= \pmatrix{0 \qquad -i \s^2 \cr -i \s^2
\qquad 0}
\qquad \qquad
\Gamma_{\underline{z}}= \pmatrix{0 \qquad -\s^3 \cr - \s^3
\qquad 0}}
\eqn\apb
$$
so that
$$
\Gamma^5 = \pmatrix{ 0 \qquad 1 \cr -1 \qquad 0} \ .
$$
With these definitions, the gamma matrices satisfy
the Clifford algebra
$$
\Gamma_{\underline{M}} \Gamma_{\underline{N}}
+ \Gamma_{\underline{N}} \Gamma_{\underline{M}}=
2 \eta_{\underline{M} \underline{N}} \ .
\eqn\apc
$$

\APPENDIX{B}{B: String Solitons and Supersymmetry}

In order to examine string and domain wall solutions,
we shall consider the following ansatz;
$$
\eqalign{
ds^2&=A^2( u, \bar u) ds^2(\bR^{1,1})+ ds^2_{(2)}
\cr
z^i&=z^i(w, \bar w)
\cr
A^a&=0}
\eqn\apd
$$
where $ds^2_{(2)}$ is a metric on the manifold
spanned by $w, \bar w$
where $w=x+iy$ and ${\bar w} = x-iy$ for $x, \ y \in \bR$.
Without loss of generality, we shall take
$$
ds^2_{(2)} = B^2(x,y) (dx^2+dy^2)
\eqn\ape
$$
to be diagonal, using the fact that any metric
on a Riemann surface is locally conformally flat.

Substituting this ansatz into the Killing spinor equations,
we find that
$$
\pd{x}A \Gamma_\ubx \e + \pd{y} A \Gamma_\uby \e + ABe^{K \over 2}
\big( {\rm Re}f + {\rm Im}f \Gamma^5 \big) \e =0 \eqn\susya
$$
together with
$$
\eqalign{
2 \pd{x} \e - \pd{x} \log A \e + \pd{y} \log {B \over A}
\Gamma_\ubx \Gamma_\uby \e
- \Gamma^5 {\rm Im} (K_i \pd{x} z^i) \e =0
\cr
2 \pd{y} \e - \pd{y} \log A \e - \pd{x} \log {B \over A}
\Gamma_\ubx \Gamma_\uby \e
- \Gamma^5 {\rm Im} (K_i \pd{y} z^i) \e =0}
\eqn\susyb
$$
and
$$
\eqalign{
\big( \rre \pd{x} z^i - \Gamma^5 \iim
\pd{x} z^i \big) \Gamma_\ubx \e
+ \big( \rre \pd{y} z^i - \Gamma^5 \iim
\pd{y} z^i \big) \Gamma_\uby \e
\cr
- Be^{K \over 2} \big( \rre (\g^{i \bar j}
D_{\bar j} {\bar f}) - \Gamma^5 \iim
 (\g^{i \bar j} D_{\bar j} {\bar f}) \big) \e =0}
 \eqn\susyc
$$

There are two cases to consider. Firstly, if the the scalar
potential vanishes, then the first Killing spinor equation
above with $f=0$ implies that $A$ is constant,
and we set $A=1$. The
remaining Killing spinor equations can be solved
by taking
$$
\eqalign{
z^i = z^i (w)
\cr
B = e^{-{K \over 2}}}
\eqn\apf
$$
i.e. the $z^i$ are holomorphic. The solution
preserves $1 \over 2$ of the supersymmetry;
$\e$ is a constant spinor satisfying
$$
\Gamma^5 \Gamma_\ubx \Gamma_\uby \e = -\e
\eqn\apg
$$

More generally, one may construct solutions for which
$f \neq 0$. In particular, we may begin by
examining \susyc\ . If we work in a real basis, so that
$$
\e = \pmatrix{ \e_1 \cr \e_2}
\eqn\aph
$$
with $\e_1$, $\e_2$ real; then \susyc\  implies
$$
\eqalign{
(\s^1 - 1) \big[ 2i \pd{w} z^i {\bar{\eta}}
+Be^{K \over 2} \g^{i \bar j} D_{\bar j} {\bar f}
\eta \big] =0
\cr
(\s^1 +1) \big[ -2i \pd{\bar w} z^i {\bar \eta}
+Be^{K \over 2} \g^{i \bar j} D_{\bar j} {\bar f}
\eta \big] =0}
\eqn\api
$$
where $\eta = \e_1 +i \e_2$. Suppose now
$\exists \  i$ such that
$\g^{i \bar j} D_{\bar j} {\bar f}=0$. Then for these $i$,
these equations may be solved by taking $z^i$ constant.
Alternatively, one may have $z^i = z^i (w)$
non-constant holomorphic with $\Gamma^5
\Gamma_{\ubx} \Gamma_{\uby} \e = - \e$;
or $z^i = z^i ({\bar w})$  non-constant anti-holomorphic
with $\Gamma^5 \Gamma_{\ubx} \Gamma_{\uby} \e =  \e$
(however if there exists more
that one value of $i$ such that
$\g^{i \bar j} D_{\bar j} {\bar f}=0$ then one cannot
have a supersymmetric solution with a mixture of
corresponding non-constant holomorphic
and anti-holomorphic complex scalars).

 Suppose now we consider $i$ for which
$\g^{i \bar j} D_{\bar j} {\bar f} \neq 0$. Define
$$
\eqalign{
\psi^i = 2i \pd{w} z^i \big[ Be^{K \over 2}
\g^{i \bar j} D_{\bar j} {\bar f} \big]^{-1}
\cr
\t^i = -2i \pd{\bar w} z^i \big[ Be^{K \over 2}
\g^{i \bar j} D_{\bar j} {\bar f} \big]^{-1}}
\eqn\apj
$$
Then one requires for these $i$;
$$
\eqalign{
(1-\s^1) \big(\psi^i {\bar \eta} + \eta \big) =0
\cr
(1+ \s^1) \big( \t^i {\bar \eta} + \eta \big) =0}
\eqn\apk
$$
There are several possibilities. Firstly, note that
one cannot have a supersymmetric solution with both
 $\psi^i=\t^i=0$. If $\psi^i=0$ then it turns out that
$\Gamma^5 \Gamma_{\ubx} \Gamma_{\uby} \e =  -\e$.
If $\t^i=0$, however, then
$\Gamma^5 \Gamma_{\ubx} \Gamma_{\uby} \e =  \e$.
Alternatively, one may have $\psi^i$, $\t^i$ both
nonzero. It turns out that if both
$|\psi^i| \neq 1$ and $|\t^i| \neq 1$ then the
 solution cannot be supersymmetric.
If however, $|\psi^i| \neq 1$ but $|\t^i|=1$ then one has
$\Gamma^5 \Gamma_{\ubx} \Gamma_{\uby} \e =  -\e$. Another
possibility is to take $|\t^i| \neq 1$ and
 $|\psi^i|=1$; then $\Gamma^5 \Gamma_{\ubx}
\Gamma_{\uby} \e =  \e$.

We shall however concentrate on the
remaining possibility, in
which we take $|\t^i|=|\psi^i|=1$ (but we do not
not necessarily
require $\t^i = \psi^i$). Writing then $\psi^i =
e^{i \theta^i}$, $\t^i = e^{i \f^i}$ for real
$\theta^i$, $\f^i$,
the supersymmetry constraint \susyc\ for $|\t^i| = 1$
 and $|\psi^i|=1$  is satisfied by taking
$$
\eqalign{
\e_1 = \sin \f^i \pmatrix{ \l^i \cr \l^i} + \sin
\theta^i \pmatrix{- \m^i \cr \m^i}
\cr
\e_2 = -(1+ \cos \f^i) \pmatrix{ \l^i \cr \l^i} -
(1+ \cos \theta^i)  \pmatrix{- \m^i \cr \m^i}}
\eqn\weird
$$

Analogous reasoning may be used to consider \susya\ .
 In particular, \susya\ may be written as
$$
\eqalign{
(\s^1 -1) \big[ 2i \pd{w} A {\bar \eta} -AB
e^{K \over 2} {\bar f} \eta \big] =0
\cr
(\s^1 +1) \big[ -2i \pd{\bar w}A {\bar \eta} -AB
e^{K \over 2} {\bar f} \eta \big] =0 \ .}
\eqn\apl
$$
Defining
$$
\eqalign{
\Omega = -2i \pd{w} A \big(ABe^{K \over 2}
{\bar f} \big)^{-1}
\cr
\Lambda = 2i \pd{\bar w} A  \big(ABe^{K \over 2}
 {\bar f} \big)^{-1}}
\eqn\apm
$$
we note that \susya\ is equivalent to
$$
\eqalign{
(\s^1 -1) (\Omega {\bar \eta} + \eta) =0
\cr
(\s^1 +1) (\Lambda {\bar \eta} + \eta) =0} \ .
\eqn\apn
$$
Hence the reasoning used to determine the various
possible values of $\psi^i$, $\t^i$
also applies to $\Omega$ and $\Lambda$.

To summarize then, neglecting the cases for which
$\Gamma^5 \Gamma_\ubx \Gamma_\uby \e = \pm \e$,
\susya\ and \susyc\ imply that
 $\Gamma^5 \Gamma_{\ubx} \Gamma_{\uby} \e \neq \pm \e$.
Furthermore, if $\g^{i \bar j} D_{\bar j} {\bar f}=0$
then $z^i$ is constant, and if $\g^{i \bar j}
D_{\bar j} {\bar f} \neq 0$ then
$\psi^i = \Omega$ and $\t^i = \Lambda$ with
$|\Omega|=|\Lambda|=1$. $\e$ is given by \weird\ .

It is also necessary to examine \susyb\ .
This constraint may be rewritten as
$$
4 \pd{\bar w} {\hat \e} -2i  \pd{\bar w} \log
{B \over A} \Gamma_\ubx \Gamma_\uby {\hat \e}
+i \Gamma^5 (-\pd{\bar w}K +2 K_i \pd{\bar w} z^i)
{\hat \e}=0 \ .
\eqn\app
$$
where ${\hat \e} = A^{-{1 \over 2}} \e$.
In this case $\t^i = \Lambda$ and $\psi^i = \Omega$
 imply (for $f \neq 0$)
$$
\eqalign{
- \pd{\bar w} z^i = A^{-1} \pd{\bar u} A {\bar f}^{-1}
 \g^{i \bar j} D_{\bar j} {\bar f}
\cr
- \pd{ w} z^i = A^{-1} \pd{ u} A {\bar f}^{-1}
\g^{i \bar j} D_{\bar j} {\bar f}}
\eqn\apq
$$
and we solve the supersymmetry constraints
by taking $\Lambda= e^{i \f}$, $\Omega= e^{i \theta}$,
for $\theta$, $\f \in \bR$ with
$$
\eqalign{
{\hat \e}_1 = \sin \f \pmatrix{{\hat \l} \cr
{\hat \l}} + \sin \theta \pmatrix{-{\hat \m} \cr {\hat \m}}
\cr
{\hat \e}_2 = -(1+\cos \f) \pmatrix{{\hat \l} \cr {\hat\l}}
-(1+\cos \theta) \pmatrix{-{\hat \m} \cr {\hat \m}}}
\eqn\apr
$$
where ${\hat \l}$, ${\hat \m} \in \bR$. Then \susyb\
implies that
$$
\eqalign{
4 \pd{\bar w} ({\hat \l} \sin \f) +i (1+\cos \f)
(\pd{\bar w}
(K+2 \log {B \over A})-2 K_i \pd{\bar w} z^i) {\hat \l} =0
\cr
-4 \pd{\bar w}((1+\cos \f) {\hat \l})
+i \sin \f (\pd{\bar w} (K+2 \log {B \over A})-
2 K_i \pd{\bar w} z^i) {\hat \l} =0
\cr
4 \pd{\bar w} ({\hat \m} \sin \theta)
+i (1+\cos \theta) (\pd{\bar w} (K-2 \log {B \over A})-
2 K_i \pd{\bar w} z^i) {\hat \m} =0
\cr
-4 \pd{\bar w}((1+\cos \theta) {\hat \m})
+i \sin \theta (\pd{\bar w} (K-2 \log {B \over A})-
2 K_i \pd{\bar w} z^i) {\hat \m} =0 \ .}
\eqn\aps
$$
This is solved by taking
$$
\eqalign{
{\hat \l} = {\xi \over \sqrt{1+ \cos \f}}
\cr
{\hat \m} = {\zeta \over \sqrt{1+ \cos \theta}}}
\eqn\apt
$$
for constant $\xi$, $\zeta \in \bR$ and $B$, $\f$ and
$\theta$ are determined by
$$
\eqalign{
\pd{\bar w} (2 \f +i (K+2 \log {B \over A}))=
2i K_i \pd{\bar w} z^i
\cr
\pd{w} (-2 \theta -i (K+2 \log {B \over A}))=
-2i K_i \pd{w} z^i}
\eqn\apu
$$
We note that these solutions generically preserve
$1 \over 2$ of the supersymmetry.

It is straightforward to check that these conditions
ensure that the
 scalar and Einstein field equations hold.

\APPENDIX{C}{C: $G_2$ Domain Walls}

\chapterstyle={\Alphabetic}
\chapternumber=3

\section{Ansatz and Killing Spinor Equations}

To find  domain wall solutions of $N=1, D=4$ supergravity,
 we shall use
the ansatz
$$
\eqalign{
ds^2&= B^2(y) [dy^2+ds^2(\bR^{1,2})]
\cr
z^i&=z^i(y)
\cr
A^a&=0 \ ,}
\eqn\dwanz
$$
where $B$ and $z^i$ will be determined by the
field equations.
Properties of domain walls in supergravity have
been reviewed in [\cveticd].
We shall also assume that they are associated with
some superpotential
$f$ which we shall not specify.
Substituting this ansatz to the Killing spinor
equations of section three, one finds
$$
\eqalign{
\pd{y} B \Gamma_\uby \e +B^2 e^{K \over 2}
\big( \rre f + \iim f \Gamma^5 \big) \e =0
\cr
2 \pd{y} \e - \Gamma^5 \iim (K_i \pd{y} z^i) \e
 +Be^{K \over 2} \Gamma_\uby (\rre f +
\iim f \Gamma^5) \e =0
\cr
\big( \rre \pd{y} z^i - \iim \pd{y} z^i \Gamma^5 \big) \e
- B e^{K \over 2} \big(
\rre (\g^{i \bar j} D_{\bar j} {\bar f}) - \iim
(\g^{i \bar j} D_{\bar j} {\bar f})
\Gamma^5 \big) \e =0 \ .}
\eqn\susdwconst
$$

We are seeking solutions that preserve two real
supercharges.
This leads to consider a set of  Killing
spinor equations which arise
as a special case of the analysis presented in
the previous appendix. In particular
for the domain wall solution we take
(for those $z^i$ directions  such that $\g^{i \bar j}
 D_{\bar j} {\bar f} \neq 0$)
$\psi^i = \t^i = \Omega = \Lambda$ with $|\Lambda|=1$.
The angles $\theta$, $\f$ defined in the appendix
 therefore satisfy $\theta = \phi$.
Then the Killing spinors are given by
$$
\eqalign{
\e_1 = {B^{1 \over 2} \sin \f \over \sqrt{1 + \cos \f}}
\pmatrix{\a \cr \b}
\cr
{\e}_2 = - B^{1 \over 2} \sqrt{1+ \cos \f}
\pmatrix{\a \cr \b}}
\eqn\dwksp
$$
for real constants $\a$, $\b$. Clearly  the
solution preserves ${1 \over 2}$ of the
 supersymmetry of $N=1, D=4$ theory.
The conditions which are implied from the Killing spinor
equations on the fields are  $\Lambda = e^{i \f}$, i.e.
$$
\pd{y} B = - e^{i \f} B^2 e^{K \over 2} {\bar f}\ ,
\eqn\dwa
$$
together with
$$
\pd{y} z^i =-B^{-1} \pd{y}B  {\bar f}^{-1}
\g^{i \bar j} D_{\bar j} {\bar f}
\eqn\dwb
$$
and
$$
\pd{y} \big[ 2 \f +iK \big] =2i K_i \pd{y} z^i \ .
\eqn\dwc
$$

\section{Dilatonic $G_2$ Domain Walls}

So far we have considered the general case of
 domain walls of $N=1,D=4$ supergravity
associated with a superpotential $f$ which
 preserve $1/2$ of supersymmetry.
Now we shall consider the special case of domain
walls in the
context of $G_2$ compactifications. In the
beginning of the analysis, we shall  keep
the superpotential $f$ arbitrary but
to give some explicit solutions of domain walls,
we shall later consider some special cases.
We recall that the K\"ahler potential is given by
$K = -{3 \over 7} \log \Theta$,
where $\Theta$ is the volume of the compact
$G_2$-manifold, and $\Theta$ is
homogeneous of degree $7 \over 3$ in $s$ coordinates
 of the moduli space $\cM$.

To proceed motivated by \lmtwo, we shall consider
solutions for which
$$
s^i = s(y) c^i
\eqn\simpdwanz
$$
for some constants $c^i$, and $p^i=0$, ie the only
 modulus field is the volume of the
compact manifold.
Then \dwc\ implies that $\f = \f_0$, where $\f_0 \in \bR$
 is constant.
Defining $f = e^{i \f_0} F$, we may write \dwa\ and \dwb\
  as
$$
\pd{y} B = - B^2 e^{K \over 2} {\bar F}
\eqn\ddwa
$$
and
$$
\pd{y} z^i = B e^{K \over 2} \g^{i \bar j} D_{\bar j}
{\bar F}\ ,
\eqn\ddwb
$$
respectively.
Furthermore because $\pd{i}K$ is homogeneous of
degree $-1$, we write
$$
K_i = \l_i s^{-1}
\eqn\dwkpt
$$
for some constants $\l_i$.
But we also know from section three that
$$
K_i = {3 \over 7} \Theta^{-1} {\partial \Theta \over
\partial s^i}.
\eqn\dwkhompt
$$
It then follows that $c^i \l_i =1$. Now \ddwb\
 may be written as
as
$$
{1 \over 2} \pd{y} K_i = B e^{K \over 2} (F K_i +
{\partial F \over \partial z^i}) \ .
\eqn\rewrconst
$$
Motivated by this we shall consider $F$ of the form
$F= F(-2 z^i \l_i)$,
so that
$$
\big({\partial F \over \partial z^i}\big)_{p=0,s^i=s c^i} =
-2 \l_i F'
\eqn\suppota
$$
where $'$ denotes differentiation with respect to $s$.
So \ddwb\ simplifies to
$$
-{1 \over 4} s^{-{1 \over 2}} \pd{y} (s^{-1}) =
 B e^{K \over 2} (s^{-{1 \over 2}}F)' \ .
\eqn\simpft
$$
In addition, because $\Theta$ is homogeneous of
 degree $7 \over 3$ in $s^\ell$
we have
$$
\Theta = \l s^{7 \over 3}
\eqn\compctvol
$$
for constant $\l$. Hence $e^{K \over 2} =
\l^{-{3 \over 14}} s^{-{1 \over 2}}$.
Eliminating $y$ from \ddwa\ and \simpft, we can
determine the component $B$ of
the metric in terms of the superpotential as
$$
B = e^{-{1 \over 4} \int {W \over s^2 W'} ds}
\eqn\onemore
$$
where $W=s^{-{1 \over 2}}F$. To find the full solution,
 it remains to substitute
\onemore\ in \simpft\ and solve for $s$. However
the resulting equation
is rather involved for a general superpotential.
To find explicit solutions additional
information is needed to describe the superpotential.

\APPENDIX{D}{D: $G_2$ and Dilatonic Black Holes}

It is convenient to make an explicit connection
between the $N=1$, $D=4$ supergravity theory with
couplings fixed by $G_2$ compactification of $D=11$
supergravity and the standard form of
Einstein-Maxwell-Dilaton supergravity such as that used in
[\shiraishi]. This supergravity theory has
 Lagrangian (in 3+1 dimensions)
$$
L= \sqrt{-g} \big[R -2 | \nabla \varphi |^2 -
e^{-2a \varphi} F^2 \big]
\eqn\enmax
$$
where $\varphi$ is the dilaton, $a$ is the constant
 dilaton coupling, $F$ is the Maxwell field strength,
and the Chern-Simons term has been neglected as we are
 considering only purely electrically (or magnetically)
charged solutions.
We recall that the portion of the $N=1$, $D=4$
supergravity action containing the curvature and
scalar terms $s^i$ is
$$
{\tilde{L}} = \sqrt{-g} \big[ R -2 k_{ij} \pd{M}
s^i \pu{M} s^j \big] \ .
\eqn\cora
$$
Suppose we consider the special case where the only
 modulus field is the volume of
the compactified $G_2$ manifold, so that $s^i = s c^i$
for constants $c^i$. Then the truncated theory
has
$$
{\tilde{L}} = \sqrt{-g} \big[ R -{1 \over 2}
 s^{-2} \pd{M} s \pu{M} s \big] \ .
\eqn\corb
$$
Then setting $s = e^{-2 \varphi}$ in \corb\ one
 obtains the curvature and scalar
portions of \enmax\ and matching the gauge field
couplings of the $N=1$, $D=4$ supergravity action with
\enmax\ we observe that the $N=1$, $D=4$ supergravity
action is equivalent to \enmax\ on taking $a=1$.

\vskip 1cm

\refout

\end